%% file: canalescarpa.tex
\documentclass[preprint,authoryear,12pt]{elsarticle}

\pdfoutput=1

\usepackage{latexsym}
\usepackage{amsmath}
\usepackage{amsfonts}
\usepackage{amssymb}
\usepackage{psfrag}
\usepackage{graphicx}

\usepackage{subfigure}
\usepackage{verbatim}
\usepackage{color}
\usepackage{soul}
\usepackage{ulem}

\journal{\texttt{xxx}}

\newcommand{\diag}[1]{ {\text{diag}\{#1\}} }

\newcommand{\R}{\mathbb{R}}

\newtheorem{theorem}{Theorem}[section]
\newtheorem{lemma}[theorem]{Lemma}

\begin{document}
\begin{frontmatter}



\title{Informative Bayesian inference for the skew-normal distribution}


\author[tony]{Antonio Canale\corref{cor}}
\author[bruno]{Bruno Scarpa}
\cortext[cor]{Corresponding author}
\address[tony]{Dept. of Economics and Statistics, University of Turin and Collegio Carlo Alberto, Italy}
\address[bruno]{Dept. of Statistical Sciences, University of Padua, Italy}

\begin{abstract}
Motivated by the analysis of the distribution of university grades, which is usually asymmetric, we discuss two informative priors for the shape parameter of the skew-normal distribution, showing that they lead to closed-form full-conditional posterior distributions, particularly useful in MCMC computation. Gibbs sampling algorithms are discussed for the joint vector of parameters, given independent prior distributions for the location and scale parameters. Simulation studies are performed to assess the performance of Gibbs samplers and to compare the choice of informative priors against a non-informative one. The method is used to analyze the grades of the basic statistics examination of the first-year undergraduate students at the School of Economics, University of Padua, Italy.
\end{abstract}


\end{frontmatter}

\section{Introduction}

The usual assumption of normality is unrealistic in many contexts, ranging from economics and finance, to medicine.
For this reason, the construction of flexible parametric distributions which allow several degrees of skewness, or kurtosis, has received increasing attention in the last two decades. 
An interesting proposal to construct such a rich class of probability distributions is given by \citet{art:azza85}, 
introducing the skew-normal family of distributions. A univariate skew-normal random variable $Y \sim SN(\xi, \omega, \alpha)$ has probability density function
\begin{equation}
	f(y; \xi, \omega, \alpha) = \frac{2}{\omega} \phi\left(\frac{y-\xi}{\omega}\right)\Phi\left(\alpha \frac{y-\xi}{\omega}\right), \, \, \, y \in \R,
\label{eq:sn}
\end{equation}
where $\xi$, $\omega$ and $\alpha$ are respectively location, scale and shape parameters, and $\phi$ and $\Phi$ are  the probability density function (pdf) and the cumulative distribution function (cdf) of a Gaussian distribution. Clearly, if $\alpha = 0$, we are back at the Gaussian distribution. A multivariate extension of the skew-normal is also available \citep{art:azzadallavalle}. A $d$-variate skew-normal random variable $Y \sim SN_d(\xi, \Omega, \alpha)$ has probability density function 

\begin{equation}
	f( {\bf y};{\boldsymbol\xi}, {\Omega}, {\boldsymbol \alpha}) = 2 \phi_d({\bf y}- {\boldsymbol\xi}; \Omega) \Phi({\boldsymbol \alpha}^T \omega^{-1}({\bf y}- {\boldsymbol\xi})).
	\label{eq:mult-sn}
\end{equation}
where ${\boldsymbol\xi}$ is a $d$-dimensional location parameter,  $\Omega$ is a $d\times d$ positive semidefinite symmetric matrix with diagonal elements $\omega^2_1,\dots,\omega^2_d$, $\omega=(\omega_1, \dots,\omega_d)^T$, and ${\boldsymbol\alpha} = (\alpha_1, \dots, \alpha_d)^T$ is a $d$ dimensional shape vector.

An interesting characteristic of the scalar skew-normal distribution is the availability of several stochastic representations which match  real world phenomena. 
For example, the scalar skew-normal distribution can be obtained by marginalization of a constrained bivariate normal distribution.
If we consider a bivariate normal distribution with non-null correlation $\delta$ and the domain of one of the components restricted to be greater than its mean, by marginalizing out this component, we obtain precisely a scalar skew-normal distribution with  shape parameter $\delta/\sqrt{1-\delta^2}$. 
This representation is particularly useful in modeling grades of university examinations taken after a placement test or in other contexts such as psychometric tests \citep{art:birnbaum50,art:arnold93}. 

Our motivating example refers to first-year undergraduate students for the program in Economics at the University of Padua (Italy). 
We want to model the distribution of these students' grades in the first class of Statistics, one of the main first-year courses. In order to be admitted to Economics, students are required to pass a preliminary placement test, the result of which is clearly correlated with the grades they will obtain for the class in Statistics.
If we assume that the joint distribution of grades for the preliminary placement test and that for Statistics are distributed as a bivariate normal distribution with correlation $\delta$, recalling the above-mentioned stochastic representation, the distribution of Statistics grades for  students admitted to the program can be viewed as skew-normal with shape parameter  $\delta/\sqrt{1-\delta^2}$.

This example also gives us extra information which may be useful. We expect that the correlation between grades on the preliminary test and the Statistics examination is positive, so we may want to use the information that the distribution of Statistics grades is skewed to the right. The Bayesian approach of inference easily allows us to include prior information about the shape parameter in the model; within this framework, we propose informative priors for this parameter.

The skew-normal class of models has been widely generalized and extended by many
authors such as \citet{art:azzadallavalle}, \citet{art:azzacapi99,art:azzacapi03}, \citet{art:brancodey} and \citet{art:gentloperf}, 
among others.
One commendable work of unification of some of the proposals  is made by \citet{arel:azza:2006}, in which the unified skew-normal (SUN) class of distribution  is introduced. If $Z \sim SUN_{m,d}(\xi,\gamma,\omega,\Omega, \Delta, \Gamma)$, the density of $Z$ is
\begin{equation}
	f(Z;\xi,\gamma,\omega,\Omega,\Delta,\Gamma) = \phi_d\left( z-\xi; \omega\Omega\omega  \right) \frac{\Phi_m(\gamma  +  \Delta \Omega^{-1}\omega^{-1}(z-\xi); \Gamma - \Delta \Omega^{-1} \Delta^T)}{\Phi_m(\gamma;\Gamma)^{-1}},
\label{eq:mainsun}
\end{equation} 
where $\Phi_d(\cdot;\Sigma)$ is the cdf of a $d$-variate Gaussian distribution with variance covariance matrix $\Sigma$, $\Omega$, $\Gamma$, and $\Omega^* = ((\Gamma, \Delta)^T, (\Delta^T, \Omega)^T)$ are correlations matrices, and $\omega$ is a $d \times d$ diagonal matrix.

Frequentist methods of inference have been proposed for estimating the parameters of the model and particular attention has been devoted to  shape parameter $\alpha$. 
Since the groundbreaking paper of \citet{art:azza85}, it has been noted that the estimation of $\alpha$ poses some intrinsic problems. Let us assume that we know that $\xi = 0$ and $\omega = 1$. In this case, the likelihood function for $\alpha$ is only the product of $n$
 standard Gaussian cdf's. If we further assume that all observations are positive (or negative), then the likelihood is monotonically increasing (or decreasing), leading to a maximum likelihood estimate of $+ (-) \infty$. In addition, even with positive and negative
observations, the profile likelihood for $\alpha$ always has a stationary point at zero and, the likelihood function may also be quite flat. 
For all these reasons, in this paper a Bayesian approach is adopted for inference on the shape parameter of the skew-normal. While focusing on the univariate case, multivariate results are also available and discussed in the following.

From an objective Bayesian viewpoint, the work of \citet{lise:lope:2006} introduces Jeffreys' reference prior, showing that it has unbounded support and is proper, if the location and scale parameters are given as known. In a more realistic scenario, in which all three parameters are unknown, in the same paper the authors discuss a closed-form expression for the integrated likelihood of $\alpha$, when location and scale parameters are integrated out. Using a similar approach, in their recent work \citet{cabras:etal} discuss default Bayesian analysis for $\alpha$,   based on suitable pseudo-likelihood and matching prior.

However, in many circumstances, as in our motivating example, prior information is available. 
Statistical literature offers some results from the subjective perspective, both in terms of nice theoretical construction and computational tractability.  This approach can be applied, reparametrizing the model and exploiting one of the stochastic representations of the skew-normal family. In this direction see for example \citet{baye:bran:2007} or \citet{arelvalle:genton:loschi:2008}  and \citet{ganc:etal:2011}. Similar results are also used by \citet{art:fruh:pyne:2010} in the context of a Bayesian skew-normal location scale shape mixture model.
A different approach is discussed by \citet{arelvalle:genton:loschi:2008}, in which the authors 
find a skew conjugate prior, given skew-normal likelihood. Their work is based on 
the shape mixture of skew-normal and  gives an  interesting point of view for Bayesian inference on the shape parameter. Despite the nice theoretical results for the posterior distribution, which turns out to be in closed form, the authors state that the proposed class of distributions is not closed under sampling of the family of distributions associated to the skew-normal likelihood and they do not discuss any tools for posterior computation.

In the next section, we discuss two priors for $\alpha$, assuming $\xi,\omega$ to be fixed {and focusing on the univariate model \eqref{eq:sn},} while showing that both lead to a closed-form full-conditional posterior distribution. Prior elicitation and a straighforward extension to the multivariate model \eqref{eq:mult-sn} are also discussed. In Section 3, exploiting one of the possible stochastic representations of the skew-normal model, we discuss an easy sampling method, particularly useful in Markov Chain Monte Carlo (MCMC) approximation of the posterior. The results are then extended to the case in which we assign an independent normal inverse-gamma prior to $\xi,\omega$, and a simulation study is presented. Section 4 compares the results of our prior with Jeffreys' non informative prior for $\alpha$. In Section 5 we analyze the data on grades in the first-year examination of Statistics by undergraduate students of the School of Economics, University of Padua, Italy, in 2011.

\section{Likelihood and prior specifications}

Let us assume that  $\xi, \omega$ are known and, without loss of generality, that $\xi=0$ and $\omega=1$. The likelihood of model (1) for an iid sample $y = (y_1, \dots, y_n)$ of size $n$ is
\begin{equation}
	f(y;\alpha) = \prod_{i=1}^n 2 \phi(y_i) \Phi(\alpha y_i).
	\label{eq:likelihood}
\end{equation}

In the following we introduce two informative prior distributions for the scalar shape parameter $\alpha$. The first is simply a normal and may be chosen in order to center the prior on a particular guess for $\alpha$. 
However, as in our motivating example about the distribution of grades of university examinations, prior beliefs are often available on the side of skewness. By addressing this case, the second proposal is itself a skew-normal distribution.
Clearly, the first proposal falls within the second one but, for the sake of discussion and clarity, we prefer to introduce and discuss the two proposals separately.

\subsection{Normal prior for $\alpha$}

We assume \textit{a priori} that the parameter $\alpha$ is normally distributed, i.e., 
\begin{equation}
	\alpha \sim \pi_1(\alpha), \,\,\, \pi_1(\alpha) = \frac{1}{\psi_0} \phi\left( \frac{\alpha -\alpha_0}{\psi_0} \right),
\end{equation}
where $\alpha_0$ and $\psi_0$ are hyperparameters reflecting prior belief about the expectation and  variance of $\alpha$.
The posterior distribution turns out to be 	
\begin{align}
	\pi(\alpha; y) &\propto \phi\left( \frac{\alpha -\alpha_0}{\psi_0} \right) \prod_{i=1}^n  \Phi(\alpha y_i) \notag \\
		& \propto \phi\left( \frac{\alpha -\alpha_0}{\psi_0} \right) \Phi_n(\alpha y;I_n) \notag \\
		& \propto \phi\left( \frac{\alpha -\alpha_0}{\psi_0} \right) \Phi_n\left(y \alpha_0  + y(\alpha -\alpha_0); I\right).
\label{eq:posterior1}
\end{align}
where $I_d$ is the identity matrix of dimension $d$.
The above equation, once normalized, belongs to the SUN class of distributions discussed in \citet{arel:azza:2006} and, more precisely,
\begin{equation}
	\alpha | y \sim SUN_{1,n}(\alpha_0 , \Delta_1 \alpha_0/\psi_0, \psi_0, 1, \Delta_1 , \Gamma_1)
\label{eq:sun1}
\end{equation}
where $\Delta_1 = [\delta_i]_{i=1,\dots,n}$ with $\delta_i = \psi_0 y_i (\psi_0^2 y_i^2 + 1)^{-1/2}$ and 
$\Gamma_1  = I - D(\Delta_1)^2 + \Delta_1 \Delta_1^T$, and where $D(V)$ is a diagonal matrix, the elements of which coincide with those of vector $V$.
Algebraic details on how to obtain such quantities are given in the Appendix.
The posterior mean and variance may be obtained from the cumulant generating function expression presented in \citet{arel:azza:2006}. Easy algebra leads to 
\begin{eqnarray*}
&&	\text{E}[\alpha;y] = \alpha_0 +  \zeta_1(\alpha_0/\psi_0 1_n; \tilde{\Gamma}),  \\
&&	\text{Var}[\alpha;y] = \psi_0^2 + \zeta_2(\alpha_0/\psi_0 1_n ;  \tilde{\Gamma}), 
\end{eqnarray*}
where $1_n$ is a $n\times 1$ vector of ones, $\zeta_{k}(x;\Sigma)$ is the $k$th derivative of $\log( 2 \Phi_n(x;\Sigma))$ with $x \in \R^n$, and
 the matrix $ \tilde{\Gamma}$ is a positive semidefinite matrix with $1/\delta_i^2$ on the diagonal and $1$ in all off-diagonal elements obtained as $\tilde{\Gamma} = D(\Delta_1)^{-1} \Gamma_i D(\Delta_1)^{-1}$.
The explicit expressions for the mean and variance of the posterior distribution are tedious to calculate and useless in practice. They do involve the calculation of $\Phi_n(x;\Gamma)$, an $n$-dimensional integral which turns out to be numerically unstable even for moderate $n$. 
Despite this apparent drawback, the above expression has a nice interpretation,  as both posterior mean and variance may be viewed as the sum of the prior expectation and variance and a data-driven quantity.

\subsection{Skew-normal prior for $\alpha$}
\label{sec:prior2}

We assume \textit{a priori} that the parameter $\alpha$ is skew-normal distributed, i.e.,
\begin{equation}
	\alpha \sim \pi_2(\alpha), \,\,\, \pi_2(\alpha) = \frac{2}{\psi_0} \phi\left( \frac{\alpha-\alpha_0}{\psi_0} \right) \Phi\left( \lambda_0 \frac{\alpha -\alpha_0}{\psi_0} \right),
	\label{eq:pi2}
\end{equation}
where $\alpha_0$ and $\psi_0$ are respectively location and scale hyperparameters
and $\lambda_0$ is a shape hyperparameter reflecting our beliefs on the direction  of  skewness.
In this case, the posterior distribution for $\alpha$ turns out to be 
\begin{align}
	\pi(\alpha; y) &\propto \phi\left( \frac{\alpha-\alpha_0}{\psi_0} \right)  \Phi\left( \lambda_0 \frac{\alpha-\alpha_0}{\psi_0} \right) \prod_{i=1}^n  \Phi(\alpha y_i) \notag \\
		& \propto \phi\left( \frac{\alpha-\alpha_0}{\psi_0} \right) 
		  \Phi_{n+1}\left( \left[ \begin{array}{c}  y\alpha_0 \\ 0 \end{array} \right] + \left[ \begin{array}{c}  y \\ \lambda_0/\psi_0 \end{array} \right] (\alpha - \alpha_0) ; I_{n+1} \right).
\label{eq:posterior2}
\end{align}

The pdf in equation \eqref{eq:posterior2} also belongs to the SUN class of distribution and, more precisely,
\begin{equation}
	\alpha|y \sim SUN_{1,n+1}(\alpha_0 , \gamma_2, \psi_0,1,  \Delta_2, \Gamma_2)
\label{eq:sun2}
\end{equation}
where 
$\Delta_2 = [\delta_i]_{i=1,\dots,n+1}$ with 
	$\delta_i = \psi_0 z_i (\psi_0^2 z_i^2 + 1)^{-1/2}$ and 
	$z = (y^T, \lambda_0\psi_0^{-1})^T$,
$\gamma_2 = (\Delta_{2;1:n}\alpha_0\psi_0^{-1}, 0)$, 
$\Gamma_2  = I - D(\Delta_2)^2 + \Delta_2 \Delta_2^T$. Equation \eqref{eq:sun2} is very close to \eqref{eq:sun1}.

An interesting case from a practical viewpoint is obtained by considering $\alpha_0 =0$. This choice for the hyperparameter is equivalent to have rough prior information only on the skewness side of the distribution of the data: indeed, assuming positive or negative values for the shape hyperparameter $\lambda_0$, puts more prior mass on the positive or negative semi-axis. We will focus on this particular case in the rest of this paper. 

The posterior mean and variance in the above case turns out to be
\begin{eqnarray*}
&&	\text{E}[\alpha;y] = \zeta_1(0_n ; \tilde{\Gamma}),  \\
&&	\text{Var}[\alpha;y] = \psi_0^2 + \zeta_2(0_n ;  \tilde{\Gamma}), 
\end{eqnarray*}
where $0_n$ is a $n \times 1$ vector of zeros and $\tilde{\Gamma}$ is defined as the in previous section. 
Similar considerations for the previous posterior distribution apply.

One could argue that the obtained posterior distributions, which are no more than special cases of those treated by \citet{arelvalle:genton:loschi:2008}, still lack practical tractability. The explicit  posterior distributions \eqref{eq:posterior1} and \eqref{eq:posterior2} in the SUN parametrization may at first seem useless and even counterproductive.  
This is not so, as the SUN parametrization allows us to build an efficient sampling method for posterior computation in MCMC, discussed in Section~3.

\subsection{Prior elicitation}
\label{elicitation}
\begin{figure}
       \centering
               \includegraphics[scale=.4]{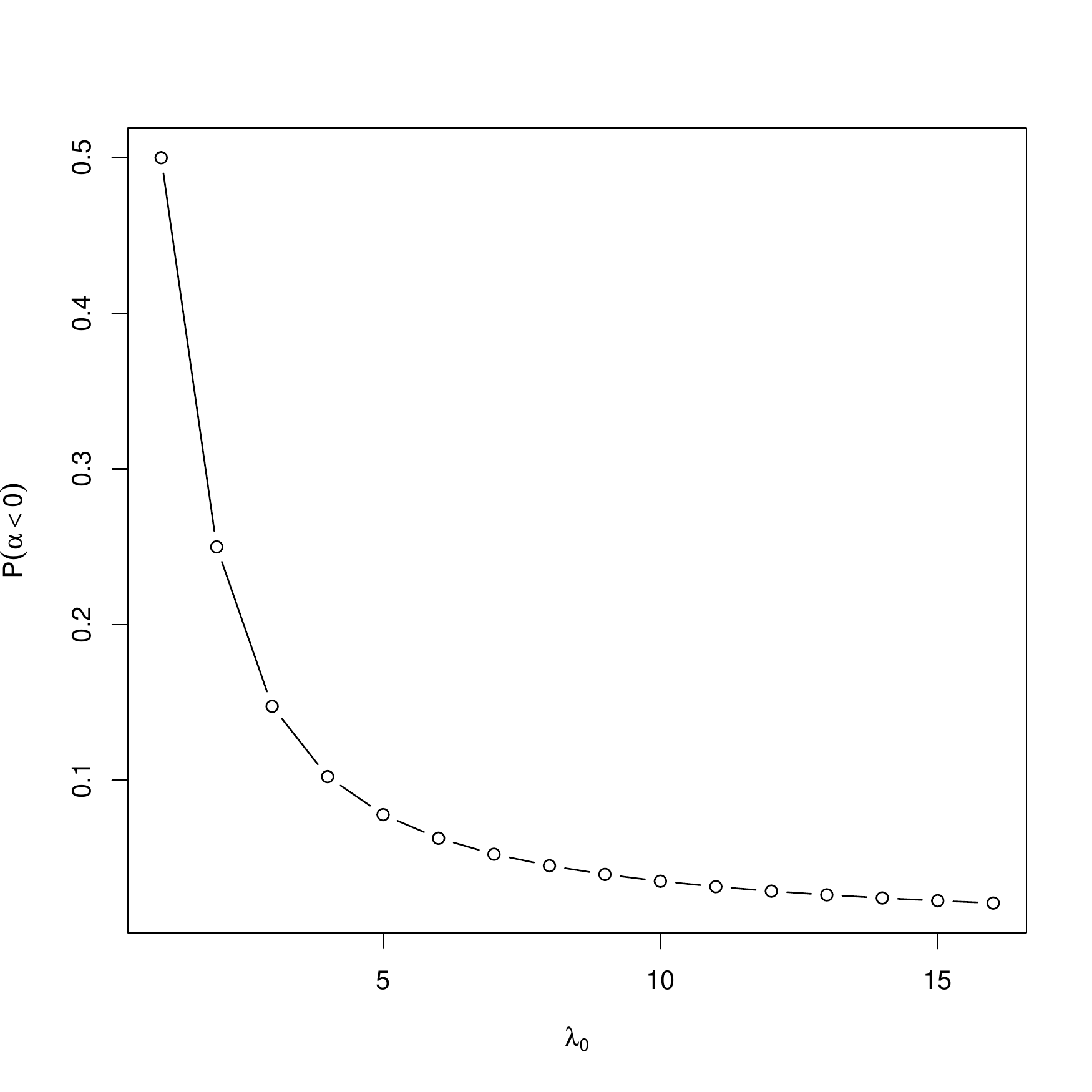}
       \caption{Probabilities mass of the occurrence of negative values of $\alpha$ for different choices of $\lambda_0$}
\label{fig:lowertail}
\end{figure}

Since we are proposing distributions useful in presence of prior information, it is of substantial interest to discuss the elicitation of the prior's hyperparameters.  
Often the sign of the skewness of the data distribution is known before analyzing data, and mild to moderate knowledge on it can be easily incorporated by using $\pi_2$ in (\ref{eq:pi2}) centered in zero. 
In this expression, a positive (negative) value of $\lambda_0$ leads to a skew prior assigning low probability mass to negative (positive) skewness. To quantify the impact of choosing $\lambda_0$ in hypothesizing the direction of skewness in this context, we plot in 
Figure~\ref{fig:lowertail} the prior probability of negative $\alpha$, $\mbox{Pr}(\alpha<0)$, for different choices of positive $\lambda_0$. 
It is evident that a very low prior mass (less than 0.05) is assumed when $\lambda_0\geq7$.
At the same time, the choice of  $\psi_0$ affects the concentration of mass around zero or on the chosen half real line.
For example, a large $\psi_0$ jointly with a high positive $\lambda_0$ corresponds to a prior belief of positive skewness but mild knowledge on the actual values of $\alpha$.

Sometimes more information on $\alpha$ are available, particularly when analysts are expert in using skew-normal distribution. 
In this case, both priors $\pi_1$ and $\pi_2$ can be used by directly centering the priors on some reasonable value and adding dispersion or even moderate skewness, according to the case.

However, often stronger prior beliefs are available on the moments of the data generating distribution.
Known relations between the parameters of the model and the first four moments allows one to incorporate these prior beliefs into the model. \citet{art:azza85} showed that
\begin{align}
& \mbox{E}[Y] 				= \sqrt{2/\pi} \delta, \notag\\
& \mbox{Var}[Y] 			= 1 - (\sqrt{2/\pi} \delta)^2,  \notag\\
& \gamma_1[Y] = \frac{4 - \pi}{2} \text{sign}(\alpha) \left[ \frac{\{\mbox{E}[Y]\}^2}{\mbox{Var}[Y]} \right]^{3/2},  \notag\\
& \gamma_2[Y] = 2 (\pi - 3) \left[ \frac{\{\mbox{E}[Y]\}^2}{\mbox{Var}[Y]} \right]^{2},
\label{eq:dptocp}
\end{align}
where $\gamma_1$ and $\gamma_2$ are the third and the fourth standardized cumulants, representing the skewness and the kurtosis of the distribution;
from these expressions, given the first four standardized cumulants, a single $\alpha$ can be obtained.
Thus, one can elicit prior hyperparameters for $\pi_1$ so that the expected skewness of the data matches the prior belief. 
The uncertainty about $\alpha$ varies according to the prior variance $\psi_0$ which can be large or small for high and low uncertainty respectively.

\subsection{Multivariate extension}

The generalization in the multivariate context is  straightforward by assuming, that the $d$ shape parameters are independent \textit{a priori}. 
Consider the multivariate likelihood arising from an iid sample ${\bf y}= ({\bf y}_1^T, \dots, {\bf y}_n^T)^T$ of size $n$ from the $d$-variate skew-normal~\eqref{eq:mult-sn} with standardized marginals and vector of means zero, namely
\begin{equation}
	f({\bf y};{\boldsymbol \alpha}) = \prod_{i=1}^n 2 \phi_d({\bf y}_i) \Phi({\boldsymbol \alpha}^T {\bf y}_i).
	\label{eq:d_likelihood}
\end{equation}

We assume that the marginal distributions of the components of ${\boldsymbol \alpha}$ are chosen between the two proposed priors; i.e. $\alpha_j \sim \pi_h(\alpha_j)$, for $h = 1,2$.
Thus, a different prior specification  may be assigned to each component of the vector $\alpha$.   
 
Say that for $d-k$ components of $ \boldsymbol \alpha$ we assume the normal prior $\pi_1$ with suitable hyperparameters and for the remaining $k$ components we assume the skew-normal prior $\pi_2$ with suitable hyperparameters. Fore sake of lightening the notation and without loss of generality, we permute the ordering of the $\bf y$ components in order to have $\alpha_j \sim \pi_2(\alpha_j)$ for the first $j = 1, \dots, k$. Hence the posterior distribution turns out to be
\begin{align}
	\pi({\boldsymbol \alpha}; {\bf y} ) 
	&\propto 
		\prod_{j=1}^d \psi_{j}^{-1} \phi\left( \frac{\alpha_j - \alpha_{0,j}}{\psi_j} \right) 
		\prod_{j=1}^k \Phi\left( \lambda_j \frac{\alpha_j - \alpha_{0,j}}{\psi_j} \right) 
		\prod_{i=1}^n \Phi({\boldsymbol  \alpha}^T {\bf y}_i) \notag \\
	&\propto 
		\phi_d\Big( \boldsymbol  \alpha - \boldsymbol  \alpha_{0}; D(\boldsymbol \psi) \Big) 
		\Phi_k\Big( \boldsymbol  \alpha_{1:k} - \boldsymbol  \alpha_{0;1:k}; D(\boldsymbol\psi\boldsymbol \lambda^{-1}) \Big) 
		\Phi_n\left( \left[ \begin{array}{c}  \boldsymbol \alpha^T {\bf y}_1 \\ \vdots \\ \boldsymbol \alpha^T {\bf y}_n \end{array} \right]; I_n \right) \notag \\
	&\propto 
		\phi_d\Big( \boldsymbol  \alpha - \boldsymbol  \alpha_{0}; D(\boldsymbol \psi) \Big) 
		\Phi_{n+k}\left(  [{\bf y}^T \boldsymbol \alpha,  D(\boldsymbol\lambda \boldsymbol \psi^{-1}) ( \boldsymbol\alpha_{1:k} - \boldsymbol  \alpha_{0;1:k}) ]; I_{n+k} \right) \notag \\ 
	&\propto 
		\phi_d\Big( \boldsymbol  \alpha - \boldsymbol  \alpha_{0}; D(\boldsymbol \psi) \Big) 
		\Phi_{n+k}\left(  
			\left[\begin{array}{c} {\bf y}^T  \boldsymbol \alpha_0 \\ \boldsymbol 0 \end{array} \right] + 
			\left[\begin{array}{c} {\bf y}^T \\ ( D(\boldsymbol\lambda \boldsymbol \psi^{-1}),\boldsymbol  0_{k\times,(d-k)}) \end{array} \right]
			(\boldsymbol \alpha - \boldsymbol  \alpha_0); I_{n+k} \right). 
\label{eq:posteriord}
\end{align}

Also the above equation, once normalized, belongs to the SUN class of distributions and, more precisely,
\begin{equation}
	{\boldsymbol \alpha} | {\bf y} \sim SUN_{d,n+k}( \boldsymbol \alpha_0 , \boldsymbol\gamma_d , D(\boldsymbol\psi^2), \Delta_d , \Gamma_d)
\label{eq:sund}
\end{equation}
where 
$\boldsymbol \alpha_0 = (\alpha_{0,1} \dots \alpha_{0,d})$ is the prior vector of locations,
$\boldsymbol \psi = (\psi_{1} \dots \psi_{d})$ is the prior vector of scales,
$\boldsymbol \lambda = (\lambda_{1} \dots \lambda_{k})$ is the prior vector of shapes,
$\Delta_d = [\boldsymbol \delta_{i}]_{i=1,\dots,n+k}$ with 
	$\boldsymbol \delta_{i} = \psi_i{\bf z}_i^T(\psi_i^2{\bf z}_i^T{\bf z}_i +1)^{-1/2}$ and 
	${\bf z} = ({\bf y}^T , ( D(\boldsymbol\lambda \boldsymbol \psi^{-1}),\boldsymbol  0_{k\times,(d-k)}) ) $ is a $(n+k) \times d$ matrix, 
$\boldsymbol\gamma_d = [(\Delta_{d}\boldsymbol \alpha_0)_{1:n},\boldsymbol 0_{k} ] $
and
$\Gamma_d = I_{n+k} - \mbox{diag}(\Delta_d \Delta_d^T) + \Delta_d D(\psi^{-2}) \Delta_d^T $ where $\mbox{diag}(A)$ is the diagonal matrix with the elements of the diagonal of $A$. 


\section{Posterior computation}

For inference on the complete vector of the parameters, we also need to specify prior distributions  for the location and  scale of the skew-normal. 
However, in this section, we first introduce an efficient algorithm to simulate the full conditional of 
$\alpha$, given $\xi$ and $\omega$. In order to describe it, we recall a useful stochastic representation 
which is the basis for our efficient Gibbs sampler.

\subsection{A stochastic representation}

The literature on the skew-normal family of distributions 
has produced abundant theoretical results on stochastic representations. 
In the following Lemma~1 we recall a particular case of one such representation of the 
SUN family, introduced in Section 2.1 of \citet{arel:azza:2006}. 
Our aim here is to exploit this theoretical result and use it as an efficient simulation algorithm for drawing observations from posterior distributions \eqref{eq:sun1}, \eqref{eq:sun2} or \eqref{eq:posteriord}.

\begin{lemma}{Lemma}{(Arellano-Valle and Azzalini, 2006)}
Let $V_0 \sim LTN_q(-\gamma; 0, \Gamma)$, $V_1 \sim N(0,1)$ with $V_0$ independent of $V_1$ and the notation $LTN_d(\tau;\mu,\Sigma)$ denotes a $d$-variate normal distribution with mean $\mu$ and variance-covariance matrix $\Sigma$ truncated at $\tau$ from below.
If
\[
	Y = \xi + \omega ( \Delta \Gamma^{-1} V_0 + \sqrt{1-\Delta^T \Gamma^{-1} \Delta} V_1  ),
\]
then $Y \sim SUN_{1,q}(\xi, \gamma, \omega, 1, \Delta, \Gamma)$.
\end{lemma}

It is evident that  simulation from the model above can be easily done relying on efficient sampling algorithms for multivariate truncated Gaussian distribution. Recent results in this direction are the slice sampler of \citet{liec:2010} or the Hamiltonian Monte Carlo of \citet{pakm:pani:2012}. 
Both these approaches require the inverse of the $n\times n$ matrix $\Gamma$ and hence their total computational cost greatly depends on computing $\Gamma^{-1}$. The computational burden increases with the sample size $n$. To perform a general matrix inversion, it is well-known that $O(n^3)$ operations are required. Given the particular expression for $\Gamma$, a closed form for its inverse is available. Using 
to the Sherman-Morrison formula \citep[e.g.,][p. 50]{golu:vanl:1989}, we can write
\begin{align*}
	\Gamma^{-1} &= \left( I - D(\Delta)^2 + \Delta \Delta^T  \right)^{-1} \\
			&=  \diag{1/(1-\delta_i^2)} - \frac{1}{1+\sum_{i=1}^n \delta^2_i(1-\delta^2_i)^{-1}} \diag{1/(1-\delta_i^2)} \Delta \Delta^T \diag{1/(1-\delta_i^2)} \\
			& = \diag{1/(1-\delta_i^2)} - \frac{1}{1+\sum_{i=1}^n \delta^2_i(1-\delta^2_i)^{-1}} \tilde{\Delta},
\end{align*}
where $\tilde{\Delta}$ is an $n\times n$ matrix with elements $\tilde{\delta}_{ij} = \delta_i \delta_j (1-\delta_i^2)^{-1} (1-\delta_j^2)^{-1} $.
 
A particular case of Lemma 1 refers to  skew-normal distribution. In this case we can simulate a skew-normal random variable $X \sim SN(\xi, \omega, \lambda)$ with its hierarchical representation in which, conditionally on $X_0$, a realization from a half normal distribution, $X$ is normal with mean $\xi + \delta X_0$
and variance $(1-\delta^2) \omega^2$. 

\subsection{A simulation study}
\label{subsec:simulation}

We checked the performance of the above prior with respect to some competitors when $n$ increases via a small simulation study, i.e. the competitors are a flat non-informative prior for $\delta = \alpha/\sqrt{\alpha^2 + 1}$, the Jeffreys' prior discussed by \citet{lise:lope:2006}, and the matching prior of \citet{cabras:etal}.
Fixing the first two parameters of the skew-normal respectively to zero and one, we have generated $10,000$ independent samples of  size $n$ = 10,  50, 100. Simulations have been run under a wide variety of true $\alpha$ parameters, leading to qualitatively similar results. We report  results for three scenarios. The first  generates samples from a skew-normal distribution with moderate skewness, a $SN(0,1,1.5)$. In the second scenario,  samples are generated from a skew-normal distribution with sharp skewness, namely $SN(0,1,-5)$. The last  generates data from a standard Gaussian distribution.

\input{tablenew}


As a posterior summary, we compute both the mean and mode of each posterior distribution. To compute the modes of all posterior distributions, we used  the function optimize of the R statistical environment \citep{R}. Estimation of the posterior mean was carried out via a Monte Carlo approximation, drawing samples of size $10,000$ from the posterior distribution and using a suitable sampling algorithm (see previous section). Calculation of the posterior mean from  Jeffreys' prior  was again done via Monte Carlo approximation with Metropolis-Hastings algorithm, rather than numerical integration, as the latter is numerically unstable (we need a first integration to compute  Jeffreys' prior and a second one to compute the posterior mean of $\alpha$). Note that the mode of the posterior distribution induced via the flat non-informative prior is equivalent to the maximum likelihood estimation.  The results, in terms of bias and mean squared error (MSE), are shown in Tables 1--3.

The posterior mode corresponding to the flat non-informative prior has clear disadvantages when $n$ is small which is reasonable. 
Indeed, in having both no prior information and little experimental information, it is hard to reach an accurate estimate. For small $n$, the posterior induced by the proposed priors, when the prior information is correct, leads to better results as expected.
With moderate $n$ and correct prior information, the final point estimates obtained with the proposed approach are always comparable to those obtained with the non-informative flat prior and the objective Jeffreys' prior. 
Instead, bias and MSE of the matching prior, maybe because of its bimodal shape, are much higher than any other approach.

However when the prior information does not match the true data generating process, the bias and MSE of the proposed point estimates are greater than those of the other approaches. 
For the second scenario, where the skewness is large, this effect is considerable even when $n$ grows. 
These results outline, once again, that a subjective approach is useful when we are really confident of our prior knowledge and in that case its usage is convenient. However, if no prior information is available, other approaches are preferable.

\subsection{An efficient Gibbs sampler for the whole parameter vector}

For inference on the complete vector of the parameters we specify an independent normal inverse gamma distribution for the location and scale parameter and the prior distributions described in previous sections for the shape parameter. Specifically we let the prior distribution for the whole vector of the parameters of model \eqref{eq:likelihood} be
\begin{equation}
	\pi(\xi,\omega,\alpha) = N(\xi; \xi_0, \kappa \omega^2) \times \text{Ga}(\omega^{-2};a, b) \times \pi_j(\alpha;\theta_0).
\end{equation}
where $\pi_j$, $j=1,2$ is one of the priors described in Section 2, with suitable hyperparameter vector $\theta_0$.

Following \citet{baye:bran:2007}, \citet{arelvalle:genton:loschi:2008} and \citet{ganc:etal:2011} and
relying on the stochastic representation of the skew-normal distribution introduced by \citet{art:azza86}, 
which, as mentioned earlier, is a particular case of Lemma 1,
we introduce independent standard normal latent variables $\eta_1, \dots, \eta_n$. 
Conditionally on such latent variables, we can consider the generic $i$-th observation as being normally distributed with mean $\xi + \delta |\eta_i|$
and variance $(1-\delta^2) \omega^2$. Thanks to this interpretation we gain conjugacy for the location and scale parameters. 
This last argument allows us to build an efficient Gibbs sampling algorithm which iterates through the following steps:
\begin{itemize}
	\item Update $\eta_i$ from its full conditional posterior distribution 
	\[	\eta_i \sim TN_{0} ( \delta (y_i - \xi), \omega^2 (1-\delta^2)) \]
	where $\delta$ is $\alpha/\sqrt{\alpha^2+1}$ and $TN_\tau(\mu, \sigma^2)$ is a mean $\mu$ variance $\sigma^2$ normal truncated below $\tau$.
	\item Sample $(\xi,\omega)$ from
	\[ 
		N\left(\hat{\mu}, \hat{\kappa} \omega^2  \right) \text{InvGam}(a + (n+1)/2, b + \hat{b})
	\]
	where
	\begin{align*}
		\hat{\mu} & = \frac{\kappa \sum_{i = 1}^n ( y_i - \delta \eta_i) + (1-\delta^2)\xi_0 }
				{n \kappa + (1-\delta^2) } \\
		\hat{\kappa} & = \frac{\kappa(1-\delta^2)}{n\kappa + (1-\delta^2)} \\
		\hat{b} & = \frac{1}{2(1-\delta^2)} \left\{ \delta^2 \sum_{i = 1}^n \eta_i^2 - 2\delta \sum_{i = 1}^n \eta_i(y_i - \xi) + \sum_{i = 1}^n (y_i - \xi)^2  + \frac{1-\delta^2}{\kappa}(\xi - \xi_0)^2 \right\}.
	\end{align*}
	\item Sample $\alpha$ from 
\[
		\alpha \sim \pi_i(\alpha|y^*) 
\]
	where $y_i^* = (y_i - \xi)/\omega$ for $i = 1, \dots, n$, and $\pi_j(\alpha|y)$, $j=1,2$ is one of the posterior distributions obtained in Section 2. 
\end{itemize}

\section{Simulation}

To assess the performance of the proposed Gibbs sampler, we analyzed simulated data in which the true values of the parameters were known. 
The data were chosen to have  behavior similar to that of the real dataset analyzed in Section 5. More precisely,  we simulate a sample of size $n=50$ from a $SN(22,3,5)$. For three different choices of prior information, we run our proposed Gibbs sampler and, after a burn-in of $2000$ iterations, we collect $10,000$ MCMC samples. 

To mimic the real data situation, an empirical Bayes approach is applied to define informative prior $\pi_1$. In the real world, data on the previous year's examinations are often known. 
Hence, we generate a different random sample of the same size and from the same distribution as the original sample, by presuming that it describes a previous year's examination results and compute the three first central moments of such a sample. As discussed in Section~\ref{elicitation} we elicit prior hyperparameters in order to match prior expectations to the previous year's sample quantities.

Such empirical information, from previous samples may be not available. 
However, given the selection mechanism noted in the introduction, we expect a positive correlation between the results of placement tests and the Statistics examinations, and thus  expect skewness to the right. Therefore, as proposed in Section~\ref{sec:prior2}, we choose as prior a skew-normal distribution with location parameter $\alpha_0 = 0$, scale parameter $\lambda_0 = 20$ and shape parameter $\psi_0 = 7$.
We expect that the average grade for the examination will be 20 or 21. 
With the already mentioned relation between central moments and direct parameterization in the skew-normal distribution, this information can be described by a normal-gamma prior for the skew-normal location and scale parameters with hyperparameters $\xi_0=21$, $\kappa = 0.25$, $a_\tau = 50$, and  $b_\tau = 250$. With these choices, we assign a prior probability of about 95\% for values of the location parameter  between 19 and 24 and about 90\% to  variance between 3 and 6. We call this prior $\pi_2$.

Convergence and mixing are diagnosed by monitoring the traceplots of the three parameters; convergence is rapid, and mixing adequate in each case, the \citet{gewe:1992} diagnostics suggesting very rapid convergence.

To compare our results with a non-informative approach within the Bayesian framework, we use Jeffreys' prior for the parameters by setting the prior probability of $(\xi, \omega)$ as proportional to $1/\omega$ and using the prior obtained by \citet{lise:lope:2006} for the shape parameters. As pointed out by the above authors, this prior for the location and scale parameters given $\alpha$ is the conditional reference prior.
To compute posterior summaries, we implement a blocked Gibbs sampler with sub-steps composed of Metropolis-Hastings steps. In this case, the burn-in is longer than for the informative proposals, and we discard the first $5,000$ iterations but still collect $10,000$ MCMC samples. Convergence and mixing are diagnosed by monitoring the traceplots of the three parameters.

\begin{figure}
\centering
\subfigure{\includegraphics[scale=.35]{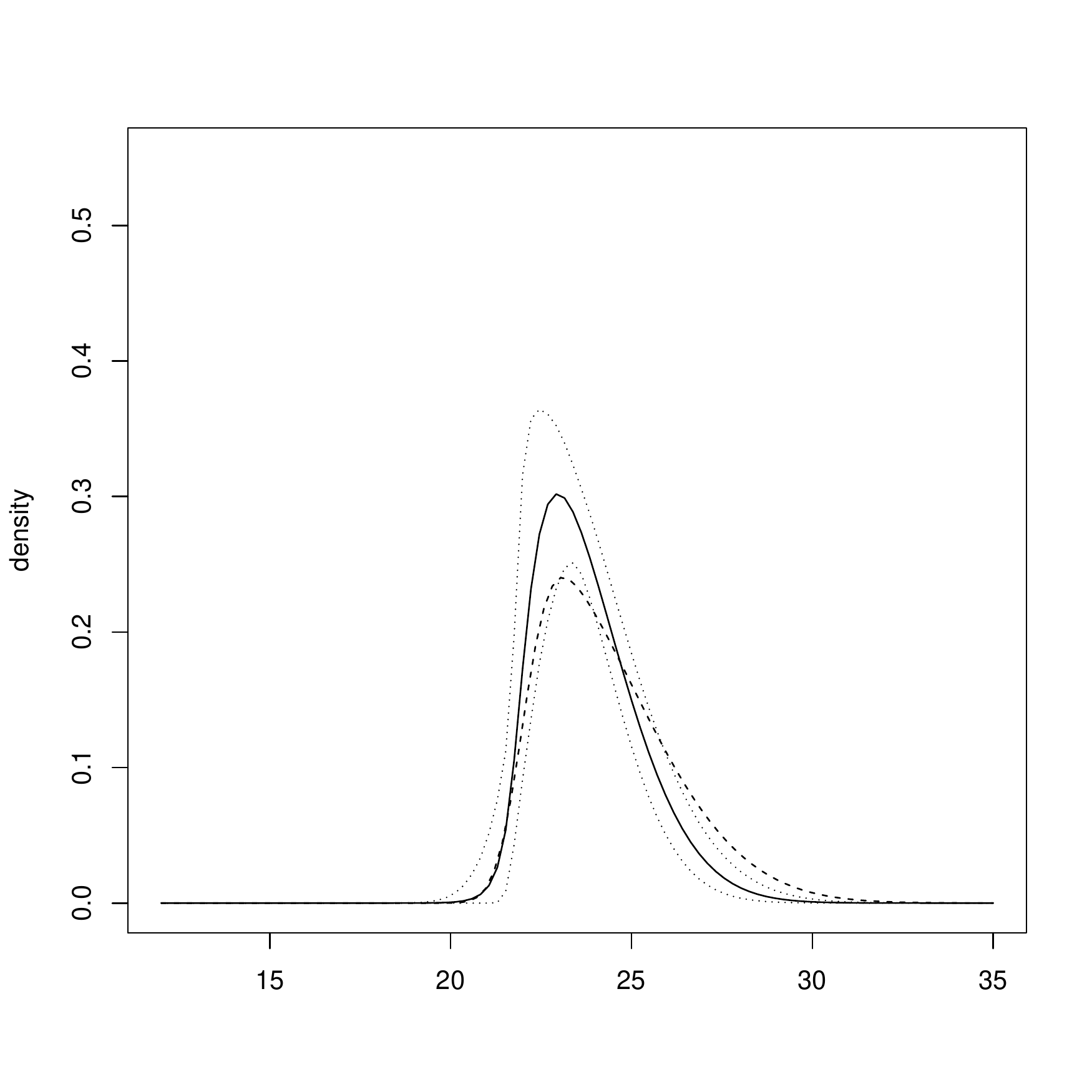}}
\subfigure{\includegraphics[scale=.35]{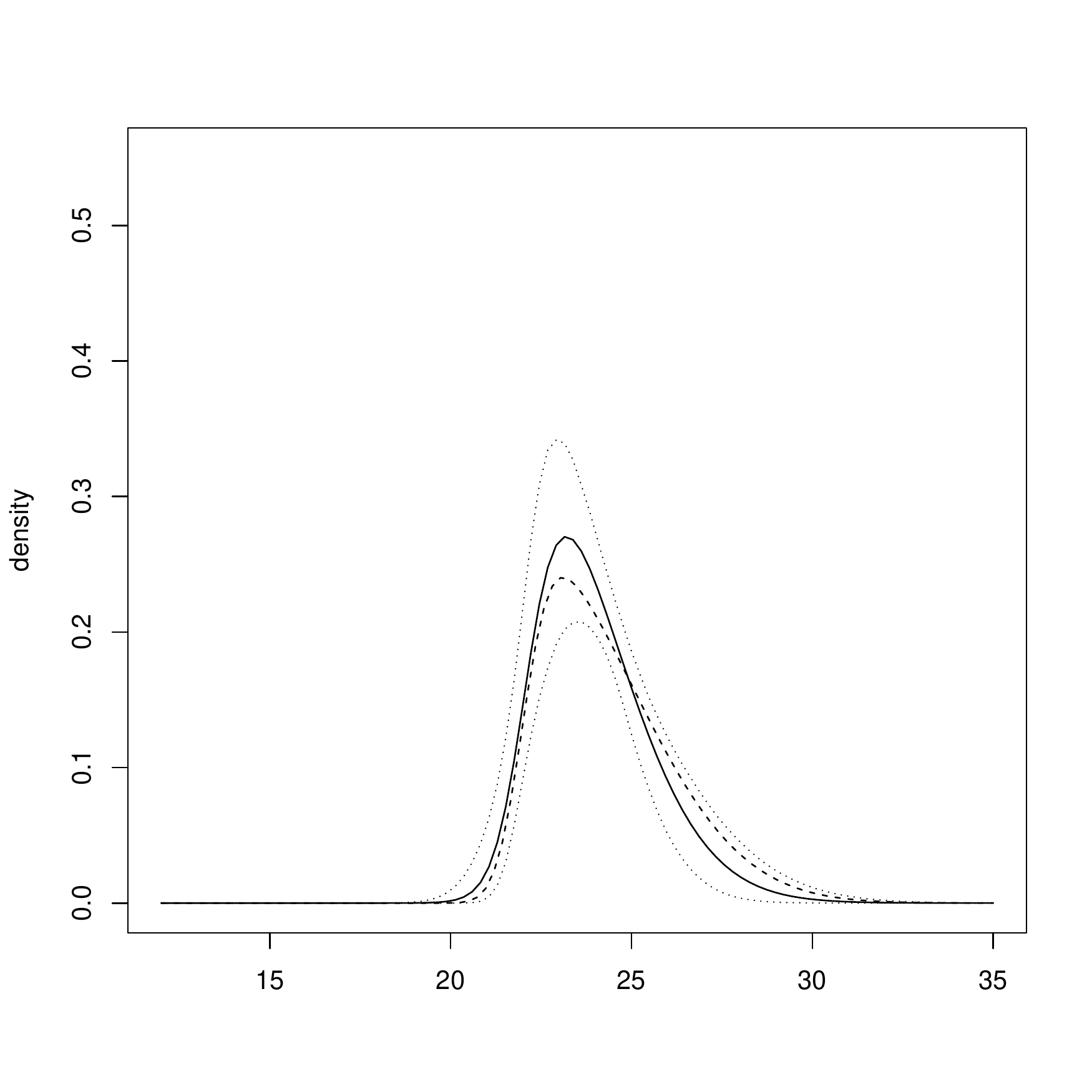}}
\subfigure{\includegraphics[scale=.35]{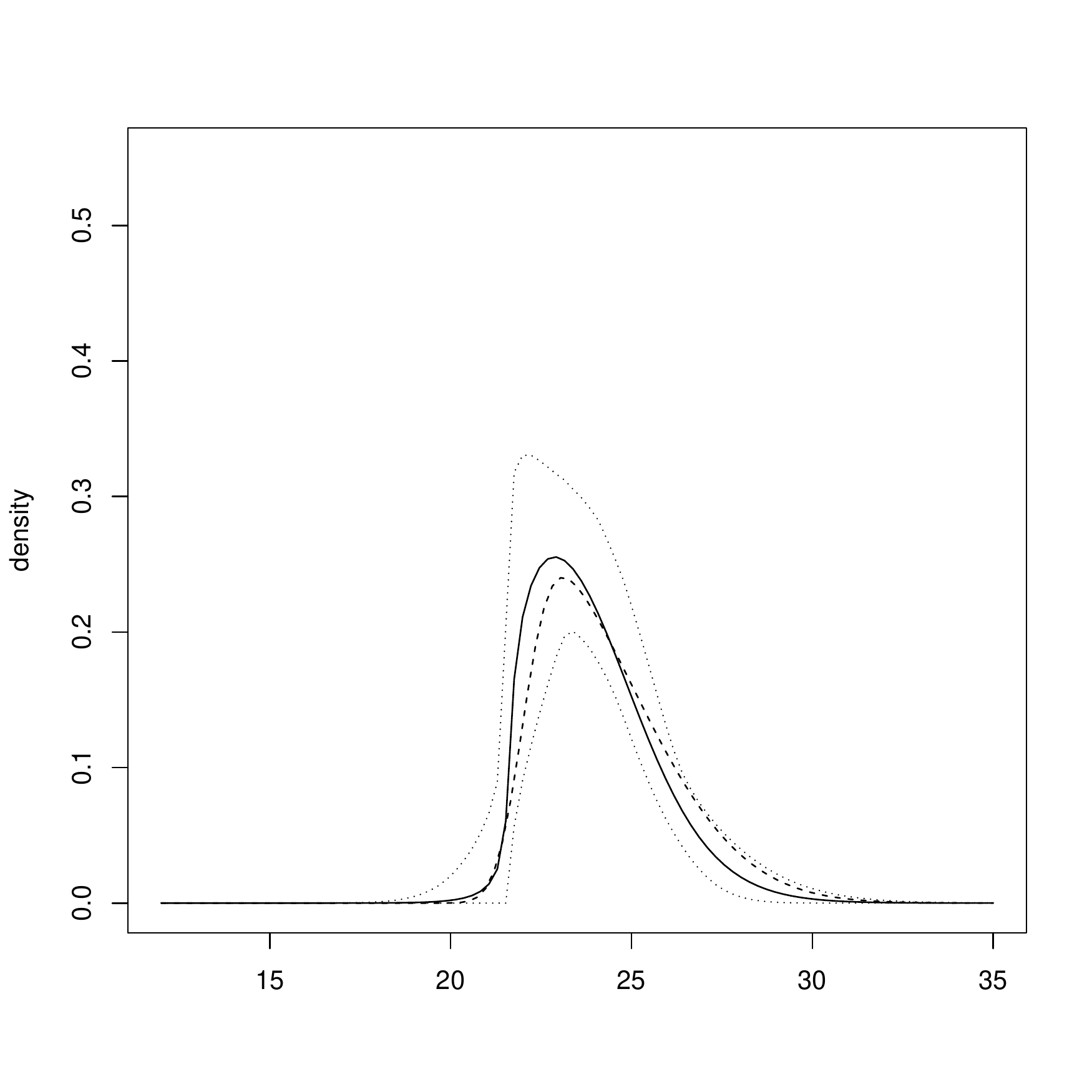}}
\caption{Posterior mean density (black line) and 95\% credible bands (dotted lines) for prior $\pi_1$ (a), prior $\pi_2$ (b) and non-informative prior (c); dashed line is true density.}
\label{fig:severaldensities}
\end{figure}

Table~\ref{tab:resgibbs} lists the posterior means and 95\% credible intervals of the parameters.
As expected, credible intervals when the non-informative prior is used are wider than the relative intervals with informative priors.
To compare the overall distributions, at each iteration we compute the value of the density function for a coarse grid of points. Figure~\ref{fig:severaldensities} reports posterior means of the densities in each sample for each case. 

\begin{table} 
\centering						
\caption{Posterior means and credible intervals for the simulated sample}
\begin{tabular}{crrr} 	\\		
\multicolumn{1}{c}{Prior} & \multicolumn{1}{c}{$\xi$} & \multicolumn{1}{c}{$\omega$} & \multicolumn{1}{c}{$\alpha$} \\ 
\hline
$\pi_1$ & 22.106 (21.765, 22.447) & 2.465 (1.917, 3.195) & 3.329 (1.976, 4.901) \\      
$\pi_2$ & 22.059 (21.727, 22.440) & 2.249 (1.965, 2.589) & 5.131 (1.936, 12.389) \\     
Jeffreys 	& 21.912 (21.450, 22.766) & 2.588 (1.869, 3.359) & 25.694 (1.318, 168.063) \\
\hline
\end{tabular}						
\label{tab:resgibbs}						
\end{table}	

\section{Application to density estimation of university grades}

We apply our informative prior to a model for estimating  the probability distribution of grades of the basic Statistics examination for first-year undergraduates of the Economics program  at the University of Padua (Italy). 
As previously mentioned, the skew-normal model seems a good model for tests taken after selection 
mechanisms, which fits the case of our data. 
Our dataset consists of grades for the 79 students which took the examination at the first session in July 2003.
%

For our inference, we first assume that our prior information is not very strong, and we only expect  that the correlation between the results of the placement test and the Statistics examination is positive. We thus choose $\alpha \sim \pi_2$ with hyperparameters $\psi_0 = 7 $, and $\lambda_0=20$. This choice is equivalent to putting less than 0.02 prior mass below zero, i.e. we strongly believe that $\alpha$ parameter is positive. This choice leads to a prior expectation for $\alpha$ of 5.58. 
 We choose the hyperparameters for normal-inverse gamma  $\xi_0 = 18$, $\kappa = 0.01$, $a =1$, and $b=5$, which lead to an expectation for $\omega$ of $1.58$. 
These choices for prior parameters correspond to assuming that \textit{a priori} data have first, second and third central standardized moments of 19.24, 0.98, and 0.88, respectively.
As a second analysis, we consider the case where data on the past year’s examination are available,
so that we can use prior $\pi_1$ while centering it in suitable quantities, as we did for the simulation study in Section 4. 
The sample mean, variance and skewness of the past year's examinations are 22.68, 13.72 and 0.35, respectively, which correspond to location, scale and shape parameters of 9.81, 18.82 and 1.67, respectively. 
We then center prior $\pi_1$ in order to have prior means matching those quantities, i.e. $\xi_0 = 9.81$, $\kappa = 0.25$, $a=1$, $b=18.82$, $\psi_0=1$, and $\alpha_0=1.67$.

\begin{figure}
\centering
\subfigure[]{\includegraphics[scale=.35]{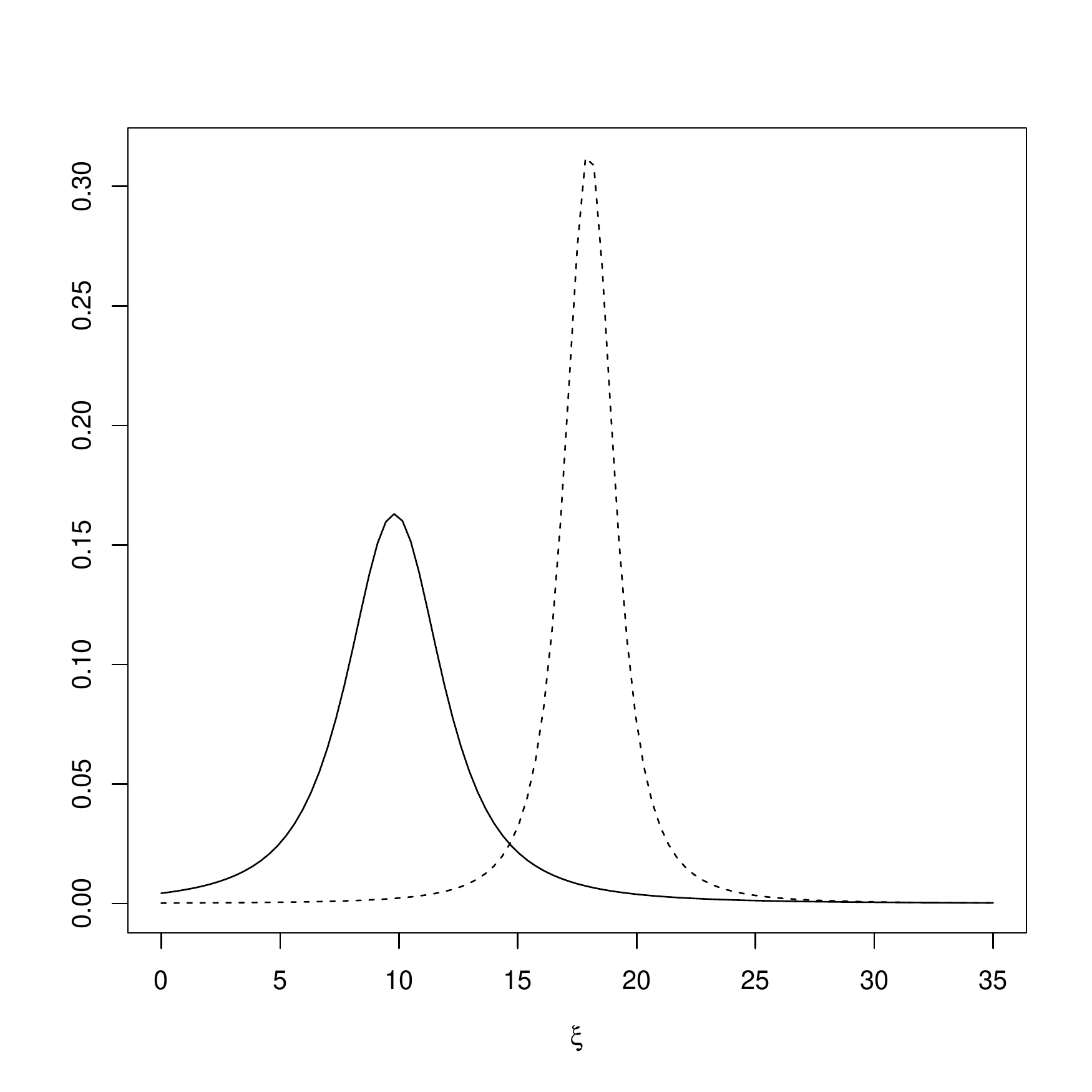}}
\subfigure[]{\includegraphics[scale=.35]{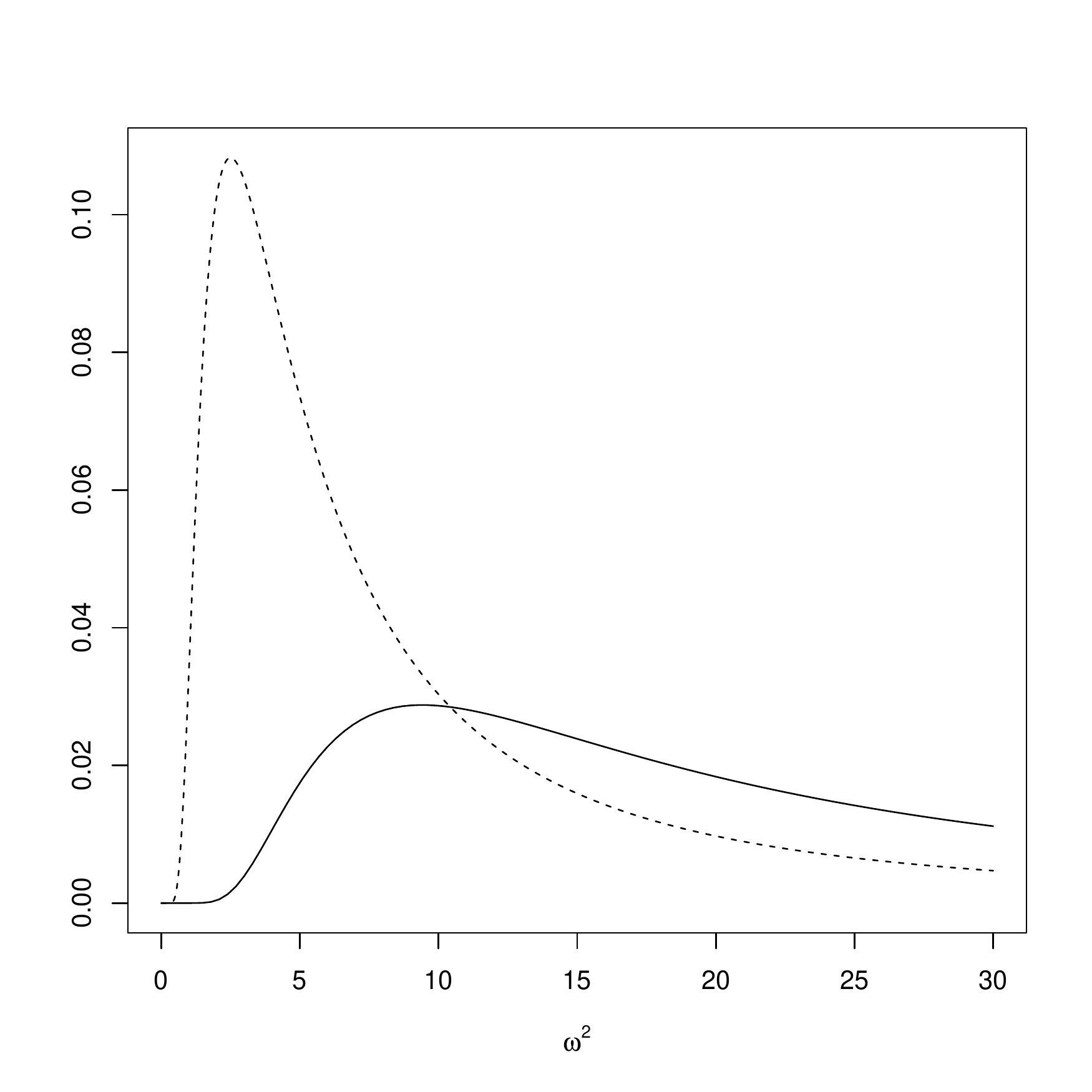}}
\subfigure[]{\includegraphics[scale=.35]{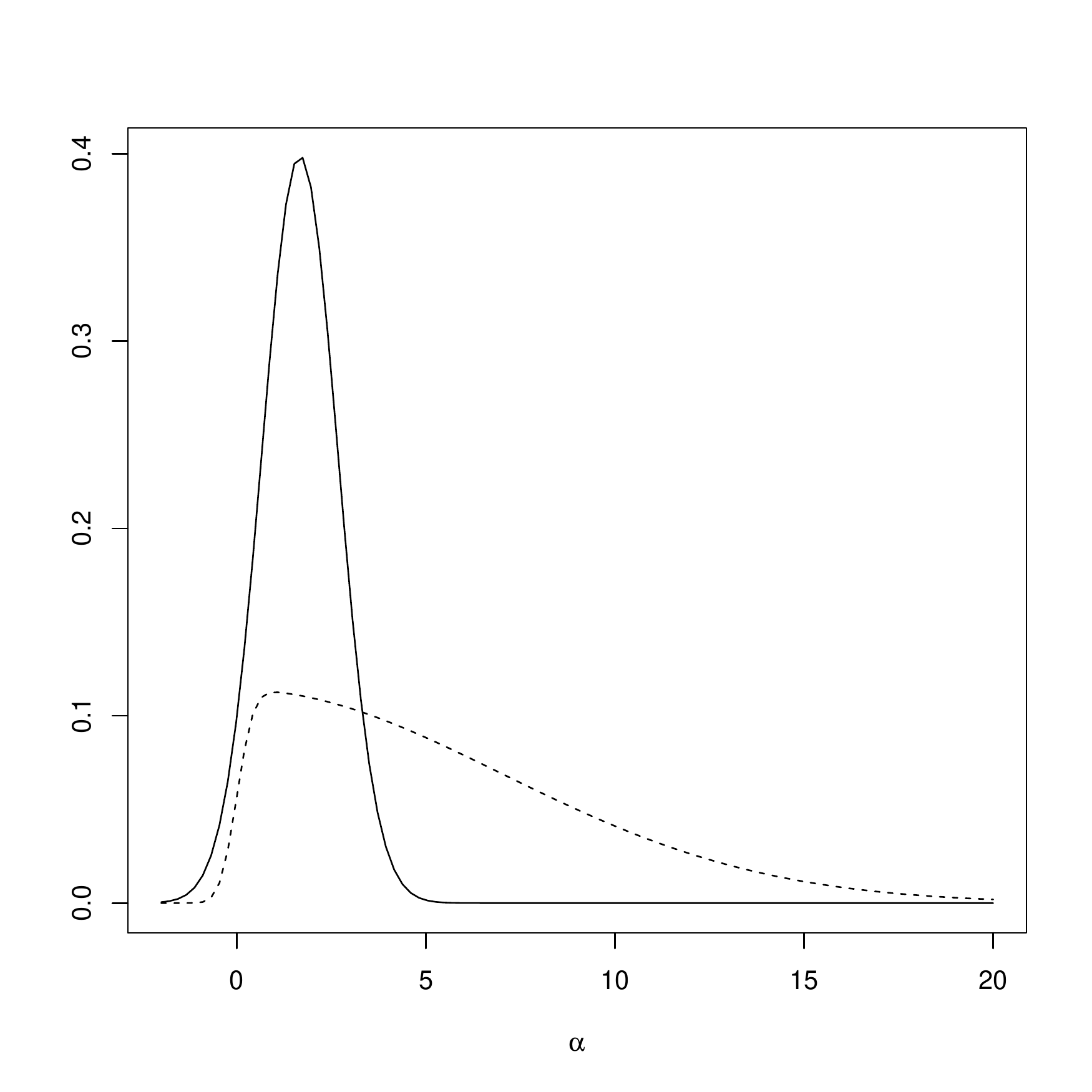}}
\caption{Marginal prior distribution with priors $\pi_1$ (continuous line) and $\pi_2$ (dashed line) for $\xi$ (a), $\omega^2$ (b) and $\alpha$ (c).}
\label{fig:priors}
\end{figure}

The resulting prior distributions are somehow different. 
In Figure~\ref{fig:priors}, the marginal  priors for the three parameters are plotted for both $\pi_1$ and $\pi_2$.
The third panel of the figure shows that the marginal prior for $\alpha$, for instance, is more concentrated around its mode, assuming $\pi_1$ rather than $\pi_2$. 
The first two panels show that the two inverse-gamma distributions are centered on very different values, leading to marginal priors for the location parameters, that is a three-parameter $t$ distribution, with different prior variability. 

\begin{figure}
\centering
\subfigure[]{\includegraphics[scale=.35]{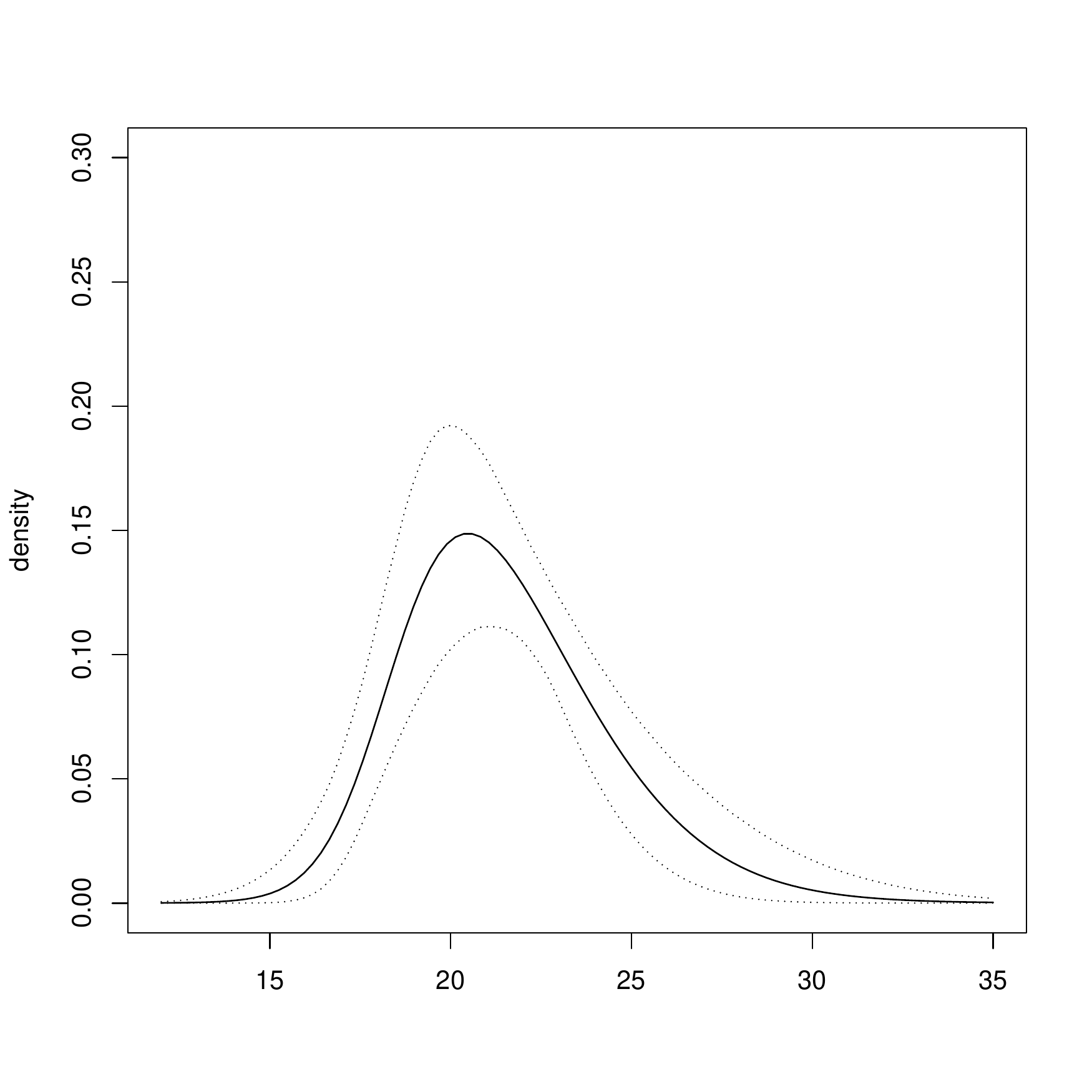}}
\subfigure[]{\includegraphics[scale=.35]{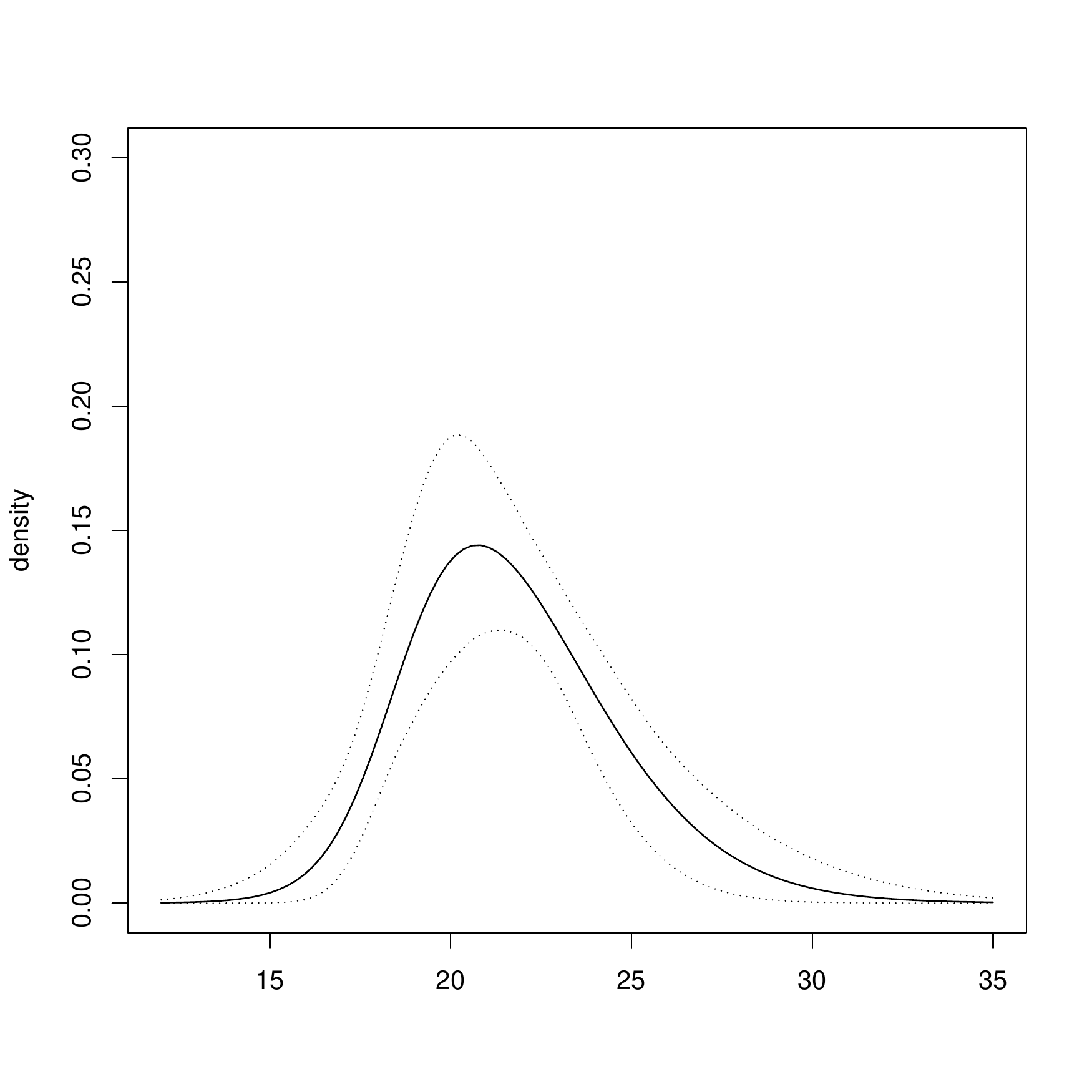}}
\caption{Posterior mean density (black lines) and 95\% credible bands (dotted lines) for prior $\pi_1$ (a) and prior $\pi_2$ (b).}
\label{fig:data}
\end{figure}

We run our Gibbs sampler for $12,000$ iterations, discarding the first 2,000 as burn-in in both cases.  The parameters values are monitored to gauge rates of apparent convergence and mixing. The traceplots of the parameters show excellent mixing and rapid convergence. Results are shown in Table~\ref{tab:realresults} and Figure~\ref{fig:data}.

Both of the final posterior densities have  modes around 21 and similar variability and skewness although the prior for the parameters were different. 
The posterior distribution obtained from prior $\pi_2$ has slightly larger posterior variability than that obtained via $\pi_1$, as shown  by the width of the credible intervals in Table~\ref{tab:realresults}. 
The use of the previous year's data to elicit hyperparameters is clearly more informative than simply assume positive skewness.

\begin{table} 
\centering						
\caption{Posterior means and credible bands for university grade dataset}
\begin{tabular}{crrr} 	\\		
\multicolumn{1}{c}{Prior} & \multicolumn{1}{c}{$\xi$} & \multicolumn{1}{c}{$\omega$} & \multicolumn{1}{c}{$\alpha$} \\ 
\hline
$\pi_1$ & 18.495 (17.688, 19.426) & 4.176 (3.125, 5.728) & 2.508 (1.224, 4.042) \\
$\pi_2$ & 18.817 (17.886, 20.229) & 4.163 (3.094, 5.767) & 2.361 (0.693, 4.556) \\        
\hline
\end{tabular}						
\label{tab:realresults}						
\end{table}	

\section{Discussion}

The literature on Bayesian inference for skew-normal distribution is mainly devoted to non-informative or objective proposals, and we find a lack of results when prior information is available, a likely situation in a variety of concrete situations. 
The main contribution of this paper is thus to discuss two informative prior distributions, with straightforward and useful multivarite extension, for the shape parameter of skew-normal distribution. Since the induced posteriors are in closed form and belong to the SUN family of distributions we described an efficient, easy and reliable sampling algorithm related to a stochastic representation of the skew-normal model which uses recent advances in sampling from multivariate truncated Gaussian distribution. Simulation studies show that with prior information, the gain in small sample size is determinant. 
A Gibbs sampling algorithm for the joint vector of the parameters is introduced and used to analyze both simulated and real data.
Also, we have confirmed that the Bayesian paradigm overcomes some limitations of the classical likelihood approach, especially when we use informative prior distributions which lead to narrower posterior credible intervals for $\alpha$ than those obtained with non-informative priors. 

\section*{Acknowledgement}
The authors thank Eric Battistin for generously providing the data.
This research was partially supported by the University of Padua CPDA097208/09 grant.

\bibliographystyle{elsarticle-harv}
\bibliography{biblio}

\section*{Appendix}

To explain the relations between equations \eqref{eq:posterior1} and \eqref{eq:sun1}, let us consider the notation in equation \eqref{eq:mainsun}. In order to match \eqref{eq:posterior1} with SUN parametrization, we set 
\begin{align*}
\xi &\leftarrow\alpha_0\\
\Psi & \leftarrow \psi_0^2\\
\gamma&\leftarrow \Delta^T \alpha_0 \psi_0^{-1}.
\end{align*}
Whit these assumptions, with $d=1$ and $m=n$, equations \eqref{eq:mainsun} becomes 
\[
 \phi_d\left( \frac{\alpha-\alpha_0}{\psi_0}  \right) \Phi_n\left(\Delta \frac{\alpha}{\psi_0}; \Gamma - \Delta \Delta^T\right),
\]
where the $n$-variate normal cdf may be rewritten as
\[
	\Phi_n\left(\Delta \frac{\alpha}{\psi_0}; \Gamma - \Delta \Delta^T\right) = \Phi_n\left( \alpha/\psi_0 1_n; \diag{1/\delta_i} (\Gamma - \Delta \Delta^T) \diag{1/\delta_i}  \right).
\]

Then we also rewrite equation \eqref{eq:posterior1} with similar steps:
\[
	\Phi_n\left(y \alpha_0  + y(\alpha -\alpha_0); I\right)  =
	\Phi_n\left(\alpha/\psi_0 1_n; \psi_0^{-2} \diag{y_i^{-2}} \right).
\]
Then, in order to obtain the parameters involved in SUN density we merely need to elicit $\Gamma$ and $\Delta$, so that $\diag{1/\delta_i} (\Gamma - \Delta \Delta^T) \diag{1/\delta_i} = \psi_0^{-2} \diag{y_i^{-2}}$.

Since $\diag{1/\delta_i}  \Delta \Delta^T \diag{1/\delta_i}$ is a $n \times n$ matrix of ones, and 
\[
 \diag{1/\delta_i} \Gamma \diag{1/\delta_i} = \left[ \frac{\gamma_{ij}}{\delta_i \delta_j} \right]_{i,j=1, \dots, n},
\]
we require $\gamma_{ij} = \delta_i \delta_j$ for the off-diagonal elements of $\Gamma$. Hence, recalling that  $\Gamma$ must be a correlation matrix, we have for each $i = 1, \dots, n$ $y_i^2 = (1-\delta_i^2)/\delta_i^2$, which defines: 
\begin{align*}
	\delta_i &\leftarrow \frac{\psi_0 y_i}{\sqrt{\psi_0^2 y_i^2 + 1}},\,\,\, \Delta = [\delta_i]_{i=1, \dots, n},\\
	\Gamma & \leftarrow I -D(\Delta)^2 + \Delta \Delta^T, 
\end{align*}
where $D(\Delta)$ is again the diagonal matrix which diagonal elements coincide with those of $\Delta$.

\end{document}

%% file: tablenew.tex
\begin{table} \centering						
\caption{Bias and mean squared error for scenario 1, ($y \sim SN(0,1,1)$)} \scriptsize						
\begin{tabular}{clrrrrrr} \\ 						
 &  &  \multicolumn{3}{c}{Bias} &  \multicolumn{3}{c}{MSE} \\ 						
 &  &  \multicolumn{1}{c}{$n=10$} &  \multicolumn{1}{c}{$n=50$} &  \multicolumn{1}{c}{$n=100$} &   \multicolumn{1}{c}{$n=10$} &  \multicolumn{1}{c}{$n=50$} &  \multicolumn{1}{c}{$n=100$}  \\ \hline						
E$_{\pi_1}$ & $\alpha_0=2, \, \psi_0=1$ &	0.624 &	0.175 &	0.058 &	0.776 &	0.136 &	0.047 \\
M$_{\pi_1}$ & $\alpha_0=2, \, \psi_0=1$ &	0.530 &	0.138 &	0.037 &	0.634 &	0.116 &	0.042 \\
E$_{\pi_1}$ & $\alpha_0=-2, \, \psi_0=1$ &	-0.564 &	-0.128 &	0.018 &	0.426 &	0.064 &	0.039 \\
M$_{\pi_1}$ & $\alpha_0=-2, \, \psi_0=1$ &	-0.625 &	-0.159 &	-0.002 &	0.479 &	0.068 &	0.037 \\
E$_{\pi_2}$ & $\lambda_0=3, \, \psi_0=1$ &	-0.013 &	0.017 &	0.007 &	0.113 &	0.067 &	0.035 \\
M$_{\pi_2}$ & $\lambda_0=3, \, \psi_0=1$ &	-0.140 &	-0.020 &	-0.012 &	0.112 &	0.061 &	0.033 \\
E$_{\pi_2}$ & $\lambda_0=-3, \, \psi_0=1$ &	-0.857 &	-0.401 &	-0.255 &	0.760 &	0.175 &	0.077 \\
M$_{\pi_2}$ & $\lambda_0=-3, \, \psi_0=1$ &	-0.848 &	-0.412 &	-0.264 &	0.741 &	0.182 &	0.081 \\
E$_U$ & &	-0.241 &	-0.062 &	-0.036 &	0.225 &	0.074 &	0.036 \\
M$_U$ & &	86.456 &	0.072 &	0.028 &	131570.542 &	0.112 &	0.042 \\
E$_{\pi_J}$ & & 	2.119 &	0.061 &	0.022 &	53.815 &	0.111 &	0.043 \\
M$_{\pi_J}$ & & 	0.033 &	0.014 &	0.002 &	0.396 &	0.088 &	0.038 \\
M$_{\text{mp}}$ & &	-0.723 &	-0.473 &	-0.326 &	3.678 &	3.015 &	1.782 \\
\hline						
\multicolumn{8}{l}{E, posterior mean; M, posterior mode; U, uniform prior for $\delta = \alpha (\alpha^2+1)^{-1/2}$;}\\ \multicolumn{5}{l}{mp, Matching prior;}						
\end{tabular}						
\label{tab:simulation1}						
\end{table}

\begin{table} \centering						
\caption{Bias and mean squared error for scenario 2, ($y \sim SN(0,1,-5)$)} \scriptsize						
\begin{tabular}{clrrrrrr} \\ 						
 &  &  \multicolumn{3}{c}{Bias} &  \multicolumn{3}{c}{MSE} \\ 						
 &  &  \multicolumn{1}{c}{$n=10$} &  \multicolumn{1}{c}{$n=50$} &  \multicolumn{1}{c}{$n=100$} &   \multicolumn{1}{c}{$n=10$} &  \multicolumn{1}{c}{$n=50$} &  \multicolumn{1}{c}{$n=100$}  \\ \hline						
E$_{\pi_1}$ & $\alpha_0=-5, \, \psi_0=2$ &	-0.224 &	-0.395 &	-0.392 &	0.454 &	0.951 &	0.973 \\
M$_{\pi_1}$ & $\alpha_0=-5, \, \psi_0=2$ &	-0.029 &	-0.172 &	-0.205 &	0.437 &	0.795 &	0.796 \\
E$_{\pi_1}$ & $\alpha_0=-15, \, \psi_0=1$ &	-9.752 &	-8.907 &	-7.974 &	95.366 &	80.296 &	65.114 \\
M$_{\pi_1}$ & $\alpha_0=-15, \, \psi_0=1$ &	-9.738 &	-8.878 &	-7.951 &	95.116 &	79.848 &	64.772 \\
E$_{\pi_2}$ & $\lambda_0=-10, \, \psi_0=3$ &	1.456 &	0.465 &	0.182 &	2.416 &	0.873 &	0.744 \\
M$_{\pi_2}$ & $\lambda_0=-10, \, \psi_0=3$ &	2.187 &	0.860 &	0.438 &	5.039 &	1.261 &	0.777 \\
E$_{\pi_2}$ & $\lambda_0=10, \, \psi_0=3$ &	4.757 &	4.091 &	3.746 &	22.638 &	16.736 &	14.037 \\
M$_{\pi_2}$ & $\lambda_0=10, \, \psi_0=3$ &	4.751 &	4.105 &	3.761 &	22.579 &	16.855 &	14.144 \\
E$_U$ & &	3.494 &	1.122 &	0.440 &	12.281 &	4.294 &	53.799 \\
M$_U$ & &	-729.715 &	-50.719 &	-0.989 &	1114827.278 &	74825.859 &	19.218 \\
E$_{\pi_J}$ & & 	-13.162 &	-3.214 &	-0.874 &	672.201 &	117.750 &	15.410 \\
M$_{\pi_J}$ & & 	2.664 &	0.028 &	-0.150 &	7.614 &	4.215 &	3.146 \\
M$_{\text{mp}}$ & &	3.583 &	0.914 &	0.266 &	15.108 &	5.407 &	3.931 \\
\hline						
\multicolumn{8}{l}{E, posterior mean; M, posterior mode; U, uniform prior for $\delta = \alpha (\alpha^2+1)^{-1/2}$;}\\ \multicolumn{5}{l}{mp, Matching prior;}						
\end{tabular}						
\label{tab:simulation1}						
\end{table}						
						
\begin{table} \centering						
\caption{Bias and mean squared error for scenario 3, ($y \sim N(0,1)$)} \scriptsize						
\begin{tabular}{clrrrrrr} \\ 						
 &  &  \multicolumn{3}{c}{Bias} &  \multicolumn{3}{c}{MSE} \\ 						
 &  &  \multicolumn{1}{c}{$n=10$} &  \multicolumn{1}{c}{$n=50$} &  \multicolumn{1}{c}{$n=100$} &   \multicolumn{1}{c}{$n=10$} &  \multicolumn{1}{c}{$n=50$} &  \multicolumn{1}{c}{$n=100$}  \\ \hline						
E$_{\pi_1}$ & $\alpha_0=0, \, \psi_0=1$ &	-0.001 &	-0.004 &	-0.004 &	0.214 &	0.033 &	0.016 \\
M$_{\pi_1}$ & $\alpha_0=0, \, \psi_0=1$ &	-0.001 &	-0.004 &	-0.003 &	0.176 &	0.031 &	0.016 \\
E$_{\pi_1}$ & $\alpha_0=10, \, \psi_0=2$ &	0.689 &	0.080 &	0.037 &	1.734 &	0.041 &	0.018 \\
M$_{\pi_1}$ & $\alpha_0=10, \, \psi_0=2$ &	0.671 &	0.079 &	0.037 &	1.634 &	0.040 &	0.017 \\
E$_{\pi_2}$ & $\lambda_0=-10, \, \psi_0=1$ &	-0.362 &	-0.139 &	-0.087 &	0.182 &	0.029 &	0.014 \\
M$_{\pi_2}$ & $\lambda_0=-10, \, \psi_0=1$ &	-0.239 &	-0.113 &	-0.076 &	0.099 &	0.021 &	0.012 \\
E$_{\pi_2}$ & $\lambda_0=10, \, \psi_0=1$ &	0.360 &	0.135 &	0.083 &	0.185 &	0.028 &	0.013 \\
M$_{\pi_2}$ & $\lambda_0=10, \, \psi_0=1$ &	0.237 &	0.109 &	0.072 &	0.101 &	0.021 &	0.011 \\
E$_U$ & &	-0.002 &	-0.003 &	-0.003 &	0.119 &	0.027 &	0.014 \\
M$_U$ & &	-0.036 &	-0.004 &	-0.007 &	1.693 &	0.041 &	0.025 \\
E$_{\pi_J}$ & & 	-0.052 &	-0.003 &	-0.003 &	2.709 &	0.033 &	0.017 \\
M$_{\pi_J}$ & & 	-0.001 &	-0.004 &	-0.003 &	0.216 &	0.031 &	0.015 \\
M$_{\text{mp}}$ & &	0.023 &	0.021 &	-0.033 &	3.308 &	3.000 &	2.088 \\
\hline						
\multicolumn{8}{l}{E, posterior mean; M, posterior mode; U, uniform prior for $\delta = \alpha (\alpha^2+1)^{-1/2}$;}\\ \multicolumn{5}{l}{mp, Matching prior;}						
\end{tabular}						
\label{tab:simulation1}						
\end{table}